\documentclass[conference]{IEEEtran}
\IEEEoverridecommandlockouts
\usepackage{cite}
\usepackage{amsmath,amssymb,amsfonts}
\usepackage{algorithmic}
\usepackage{graphicx}
\usepackage{textcomp}
\usepackage{nicefrac}

\usepackage{tabularx}
\usepackage{enumitem}

\usepackage{xcolor}

\def\BibTeX{{\rm B\kern-.05em{\sc i\kern-.025em b}\kern-.08em
    T\kern-.1667em\lower.7ex\hbox{E}\kern-.125emX}}

\usepackage{tikz}
\usepackage{textcomp}
\usepackage{hyperref}

\newcommand\copyrighttext{%
	\footnotesize \textcopyright 2024 IEEE. Personal use of this material is permitted.
	Permission from IEEE must be obtained for all other uses, in any current or future
	media, including reprinting/republishing this material for advertising or promotional
	purposes, creating new collective works, for resale or redistribution to servers or
	lists, or reuse of any copyrighted component of this work in other works.}
\newcommand\copyrightnotice{%
	\begin{tikzpicture}[remember picture,overlay]
	\node[anchor=south,yshift=10pt] at (current page.south) {\fbox{\parbox{\dimexpr\textwidth-\fboxsep-\fboxrule\relax}{\copyrighttext}}};
	\end{tikzpicture}%
}

\begin{document}
%Performance Evaluation of Dedicated Restricted Target Wake Time Usage for Real-Time Applications
\title{Dedicated Restricted Target Wake Time for Real-Time Applications in Wi-Fi 7\\
\thanks{This research was partially funded by the
ICON project VELOCe (VErifiable, LOw-latency audio Communication), realized in collaboration with imec, with project support from VLAIO (Flanders Innovation and Entrepreneurship).}
}

\author{\IEEEauthorblockN{
		Andrey Belogaev\IEEEauthorrefmark{1},
		Xiaoman Shen\IEEEauthorrefmark{2},
		Chun Pan\IEEEauthorrefmark{2},
		Xingfeng Jiang\IEEEauthorrefmark{2},
		Chris Blondia\IEEEauthorrefmark{1}, and
		Jeroen Famaey\IEEEauthorrefmark{1}}
% <-this % stops an unwanted space
\IEEEauthorblockA{\IEEEauthorrefmark{1}IDLab, University of Antwerp -- imec, Belgium\\
	Email: \{andrei.belogaev, chris.blondia, jeroen.famaey\}@uantwerpen.be}
\IEEEauthorblockA{\IEEEauthorrefmark{2}Huawei Technologies Co., Ltd., P.R. China
\\
	Email: \{shenxiaoman, panchun, jiangxingfeng\}@huawei.com}
}

\maketitle
\copyrightnotice

\begin{abstract}
Real-time applications (RTA) tend to play a crucial role in people's everyday life. Such applications are among the key use cases for the next generations of wireless technologies. RTA applications are characterized by strict guaranteed delay requirements (in the order of a few milliseconds). One of the pillars of enabling RTA in next-generation Wi-Fi standards is Restricted Target Wake Time (R-TWT), which provides Wi-Fi stations exclusive channel access within negotiated service periods (SPs). If each RTA data flow uses dedicated SPs for data transmission, they are completely isolated from each other and do not experience any contention. To ensure the satisfaction of RTA QoS requirements while minimizing the channel airtime consumption, it is important to properly select the R-TWT parameters, namely the duration of SPs and the period between SPs. In this paper, we develop a mathematical model that estimates the delay probability distribution and packet loss probability for a given set of network, traffic and R-TWT parameters. Using this model, the access point can select the optimal R-TWT parameters for the given QoS requirements. The high accuracy of the model is proven by means of simulation.
\end{abstract}


\vspace{-0.6cm}
\section{Introduction}
\vspace{-0.1cm}

Wireless networks are continuously evolving, following the ever-increasing user demands for a variety of Quality of Service (QoS) performance indicators such as data rate, delay, jitter, packet loss ratio, energy efficiency, coverage, and many others. One of the key targets for the upcoming generations of Wi-Fi networks are real-time applications (RTA), which are steadily becoming a significant part of our lives. In particular, applications such as AR/VR, industrial automation, robotics and real-time gaming require strong guarantees in terms of delay, which should be no more than a few milliseconds in the wireless network segment~\cite{rta2018report}. To tackle these new challenges, the IEEE Task Group BE (TGbe) initiated a new amendment to the Wi-Fi standard (IEEE 802.11be) in 2019, which is currently under development~\cite{khorov2020current}. This amendment introduces many novel features to improve Wi-Fi performance in different directions, e.g., nominal throughput, reliability of data delivery, spectrum efficiency, delay guarantees. One of the key features introduced in Wi-Fi 7 for the latter is Restricted Target Wake Time (R-TWT), which aids delay-sensitive data flows to obtain a contention-free channel resource~\cite{ieee80211be}.

Initially, the Target Wake Time (TWT) mechanism was introduced in IEEE 802.11ah~\cite{tian2021wi} and inherited by IEEE 802.11ax~\cite{nurchis2019target} to provide low energy consumption to energy-limited devices. This mechanism allows two Wi-Fi devices, usually an access point (AP) and a station (STA), to establish a TWT agreement and negotiate periods of time called Service Periods (SP) where they will exchange data. Outside SPs, the STA can save its energy by remaining in the doze state. Besides the reduction of the energy consumption, this mechanism also impacts the delay. By proper assignment of TWT SPs to different STAs, the contention in those SPs can be reduced, since the stations in doze state do not contend for the channel. To further protect the members of a TWT agreement from contention, IEEE 802.11be proposes R-TWT. The word ``restricted'' in R-TWT means that only the members of the TWT agreement can transmit data within an R-TWT SP, while all other STAs shall finish their transmissions prior to the start of this SP. 

The novel R-TWT mechanism opens up new opportunities for differentiated QoS provisioning in Wi-Fi networks. It has been proved that the basic traffic differentiation approach using access categories (AC) does not provide strong guarantees on QoS performance indicators, e.g., required for industry automation scenarios~\cite{memon2021survey}. Specifically, the approach is not flexible enough, because the standard offers only four access categories (voice, video, best effort and background), for each of which a device supports different hardware queues with individual parameters of random channel access. To overcome the limitations of this approach, the idea of airtime slicing was proposed~\cite{isolani2021support, richart2020slicing}, which is based on allocating different airtime shares to data flows according to their QoS requirements. However, the amount of airtime for each share is decided by a slice orchestrator, which is implemented in software. Thus, to provide strong guarantees, tight interaction between hardware queues and the software orchestrator is required. Besides, packets experience additional latencies when they wait in software queues before the airtime is granted. There are solutions in literature that are developed specifically for RTA applications, e.g.,~\cite{chemrov2022smart, xia2023hybrid}, and do not require software queues to control the resource sharing. However, in contrast to R-TWT, they are not considered for standardization. The flexibility provided by R-TWT allows directly allocating time slots to hardware queues and even for certain traffic identifiers. Moreover, it allows completely isolating data flows by allocating them dedicated time slots, so that the data transmission is not interrupted by other flows.

The R-TWT parameters should be selected so that the QoS requirements are satisfied with a given probability, while consuming minimum channel time. We focus on RTA applications that require maintaining delays of a few milliseconds with very high probability, e.g., $99.9\%$. For selection of proper R-TWT parameters, analytical instruments are needed that allow estimating the QoS performance indicators of the data flow. Specifically, the accurate estimation of delay distribution is important for the prediction of crucial RTA performance indicators, such as average delay, jitter and delay percentiles.

There are multiple works in literature that study the network performance of TWT or similar mechanisms. In papers~\cite{yang2018energy, santi2019accurate, stepanova2020joint}, performance of the TWT mechanism is investigated. Since TWT was proposed mainly for power management, the focus is on the energy efficiency, while the delay is left out of scope. Works~\cite{guo2020performance, khorov2020modeling} investigate performance of a network that uses the Time Division Multiple Access (TDMA) based mechanisms. In particular, Guo et al.~\cite{guo2020performance} study the coexistence between TDMA and random channel access. However, the traffic served using the TDMA schedule is not isolated, so STAs have to defer their transmissions until the end of other transmissions initiated by random channel access. Khorov et al.~\cite{khorov2020modeling} propose a mathematical framework for channel access in periodic reserved time intervals. However, this framework relies on the assumption that packets are dropped when their delay reaches a given threshold, which leads to delay underestimation. Besides, the models in both papers do not allow estimating the delay distribution or percentile.  

To the best of our knowledge, there are no papers in literature devoted to the analytical modeling of the R-TWT mechanism, or more precisely, to the estimation of packet delay distribution for RTA applications. In this paper, we propose a light-weight yet accurate approach for estimating the packet delay distribution and loss probability for a given set of traffic, network and R-TWT parameters. The numerical results show that the accuracy of the model makes it applicable to R-TWT parameter selection for the given RTA QoS requirements.

The rest of the paper is organized as follows. 
In Section~\ref{sec:system_model}, we describe the considered system and formulate the problem statement.
In Section~\ref{sec:math_model}, we develop the analytical model aimed at estimating the network performance indicators.
In Section~\ref{sec:numerical_results}, we validate the analytical model by means of simulation and analyze numerical results obtained with it.
Finally, in Section~\ref{sec:conclusion} we conclude the paper.

\vspace{-0.15cm}
\section{System model}
\label{sec:system_model}
\vspace{-0.1cm}

Each R-TWT agreement between a STA and an AP can be described with the following parameters:
\begin{itemize}
	\item \textit{R-TWT offset}, which indicates the position of the first SP;
	\item \textit{R-TWT SP duration}, which is the amount of time available for data exchange;
	\item \textit{R-TWT period}, or wake interval, which is the duration of the time interval between two consecutive R-TWT SPs.
\end{itemize}

We consider RTA data flows that use dedicated R-TWT SPs for data transmission. Then, each data flow can be considered separately, because they do not interact with each other. Since there is no contention inside R-TWT SPs, we completely disable the random access procedure.

Denote the duration of R-TWT SP as $SP$ and R-TWT period as $T$ (for convenience, we list all the notations in Table~\ref{table:notation}). We assume that RTA flows transmit data only inside R-TWT SPs. We call the interval between two consecutive R-TWT SPs a vacation, whose duration equals $V = T - SP$. Each RTA flow is modeled as a Poisson flow with intensity $\lambda$ packets per time unit and packet transmission duration $S$ time units (including the time required for acknowledgment transmission in the opposite direction). The R-TWT SP duration is selected so that it allows transmitting $N$ packets, i.e., $SP = N \cdot S$.

\begin{figure}[t]
	\centerline{\includegraphics[width=0.8\linewidth]{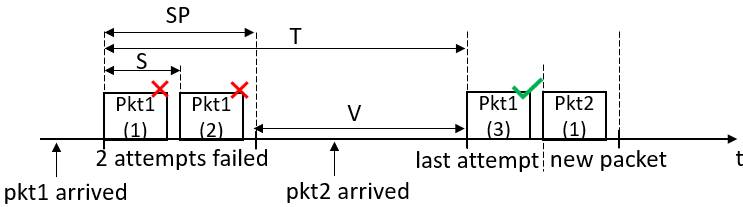}}
	\vspace{-0.2cm}
	\caption{System model for $N = 2, R = 3$.}
	\vspace{-0.6cm}
	\label{fig:system_model}
\end{figure}

Since the packets are transmitted over an error-prone wireless channel, their transmissions can fail due to channel errors. Denote the packet error probability $p_{err}$. For each packet, we allow at most $R$ transmission attempts. If all $R$ attempts fail, then the packet is considered lost (cf., Figure~\ref{fig:system_model}).

During the R-TWT SP it might happen that the station does not have any data to transmit. In this case, we assume that it does not finish its R-TWT SP earlier. Then, if a packet arrives and there is still enough time for at least one transmission attempt, the station can proceed with the transmission.

The AP has to select the optimal R-TWT parameters that guarantee QoS requirements for an RTA application while minimizing the airtime consumption. In our case, minimization of the airtime consumption is equivalent to maximization of the \textit{system capacity} $C = \frac{T}{SP}$, which is the maximum number of RTA data flows that can be served simultaneously. For optimal parameter selection, the AP needs to estimate the performance of the flow for the given R-TWT parameters. Hence, in this paper we develop an analytical model, that allows estimating the packet delay distribution and the packet loss probability for the given set of parameters: 1) traffic parameters $\{S, \lambda\}$; 2) network parameters $\{p_{err}, R\}$; 3) R-TWT parameters $\{SP, T\}$. Using the delay distribution, we can calculate performance indicators crucial for RTA, such as average delay, jitter and delay percentiles.

\vspace{-0.1cm}
\section{Analytical model}
\label{sec:math_model}
\vspace{-0.1cm}

\subsection{Notations}

We divide the timeline into slots where the length of a slot equals $S$. Let the duration of vacation equal $M = \frac{V}{S}$ slots. For continuous-time M/G/1 systems similar to the considered system, a classical approach to obtain the delay distribution is to calculate its Laplace transform~\cite{kleinrock1975queueing, lee1989m}, while the actual distribution can be obtained via the inverse Laplace transform. Although the inversion procedure for simple functions can be done analytically, for complex functions it often requires numerical approaches, such as De Hoog's method~\cite{de1982improved}, which have high computational complexity. In contrast, by approximating the continuous-time system with a slotted batch-arrival system, we avoid the usage of Laplace transform, and provide a fast yet accurate approach to estimate the delay distribution. As such, we model multiple transmission attempts by replacement of a packet arrival with a batch of packets, whose size $1 \leq r \leq R$ equals the number of transmissions required for a successful delivery of the packet.

We assume that $S \ll \frac{1}{\lambda}$, i.e., the slot duration is much shorter than an average interval between packet arrivals. For example, according to~\cite{rta2018report}, typical packet size for RTA applications is several hundreds of bytes (e.g., $100-200$ bytes for console gaming, $30-300$ bytes for industry automation) and typical traffic intensity is of the order of $60$ Hz (i.e., $\sim 16$ ms interval between packets). The transmission duration for a packet of $200$ bytes in a $20$ MHz channel on the lowest modulation-and-coding scheme (MCS) including the acknowledgment does not exceed $300$\textmu s, which is more than $50$ times smaller than the $16$ ms inter-packet interval. Then, we further assume that only one batch can arrive during a slot and it can happen only in the beginning of the slot. Since the number of packets that arrive during a certain time interval $S$ has Poisson distribution $p(l) = \nicefrac{(\lambda S)^l e^{-\lambda S}}{l!}$, the probability that zero batches arrive equals $b_{0} = p(0) = e^{-\lambda S}$.

Then, the remaining probability corresponds to a batch arrival during a slot, and equals $b = 1 - e^{-\lambda S}$.

By definition, the length of each batch has a truncated geometric distribution with failure probability $p_{err}$ and maximum size $R$, i.e., the probability that a batch has size $1 \leq r < R$ equals $p_{err}^{r-1} (1 - p_{err})$. If the batch size is less than $R$, then the packet is always successfully transmitted. For batches of size $R$ two cases are possible: 1) if the last transmission attempt is successful, then the packet is successfully transmitted; 2) otherwise, the packet is lost.

Let us call a batch \textit{successful} if its transmission results in a successful packet delivery. Otherwise, we call the batch \textit{failed}. The probability that a successful batch of size $1 \leq r \leq R$ arrives in the slot equals $b_r = b (1-p_{err}) p_{err}^{r-1}$.

Similarly, the probability that a failed batch of size $R$ arrives in the slot, i.e., packet loss probability, equals
\vspace{-0.11cm}
\begin{equation}
b_{fail} = b \cdot p_{err}^{R} \equiv b \left[ 1 - \sum\limits_{r=1}^{R} (1-p_{err}) p_{err}^{r-1} \right]
\end{equation}

To limit the number of system states to a finite number, we limit the queue size to $K$ packets, including the one being transmitted. We assume that a whole batch is dropped if it cannot be stored as a whole in the queue. We select $K$ big enough to make the probability of overflow negligible. Hence, we consider the dropped batches neither failed nor successful.

Although the delay calculation for the failed batches is meaningless, we still spend time on their transmissions. Thus, we also introduce probabilities that a batch of size $r$ arrives in the slot regardless it is successful or not:
\begin{eqnarray}
\hat{b}_{r} = b_{r}, \; \forall \; r \in [0, R-1]; \; \hat{b}_{R} = b_{R} + b_{fail}
\end{eqnarray}

\begin{table}[t]
	\centering
	\caption{\label{table:notation} List of used variables.}
	\vspace{-0.2cm}
	\setlength{\tabcolsep}{3pt}
	\begin{tabularx}{\columnwidth}{|p{0.04\textwidth}|X|}
		\hline
		$SP$ & Duration of R-TWT SP \\
		$T$ & R-TWT period \\
		$V$ & Duration of vacation \\
		$\lambda$ & Intensity of the data flow \\
		$p_{err}$ & Probability of failure due to channel errors\\
		$S$ & Packet transmission duration, i.e., slot duration \\
		$N$ & Duration of R-TWT SP in slots \\
		$R$ & Number of allowed transmission attempts \\
		$M$ & Duration of vacation in slots \\
		$C$ & System capacity \\
		$b_0$ & Probability that zero batches arrive in the slot \\
		$b$ & Probability that one batch arrives in the slot \\
		$b_r$ & Probability that a successful batch of size $r$ arrives in the slot \\
		$b_{fail}$ & Probability that a failed batch of size $R$ arrives in the slot \\
		$\hat{b}_{r}$ & Probability that a batch of size $r$ arrives in the slot \\
		$K$ & Maximum number of packets that can be stored in queue \\
		$k$ & Number of packets in the queue \\
		$n$ & Slot sequence number \\
		$p_{k,n}$ & Stationary distribution of Markov chain \\
		$\tilde{p}^{k,n}_{r}$ & Probability of $r$-packets batch arrival in slot $n$ with $k$-packets queue\\
		$D_{r}^{k,n}$ & Delay for a $r$-packets batch arrived in slot $n$ with $k$-packets queue\\
		$D$ & Packet delay\\
		\hline
	\end{tabularx}
	\vspace{-0.3cm}
\end{table}

\vspace{-0.2cm}
\subsection{Markov chain}
\vspace{-0.1cm}

\begin{figure}[t]
	\centerline{\includegraphics[width=0.9\linewidth]{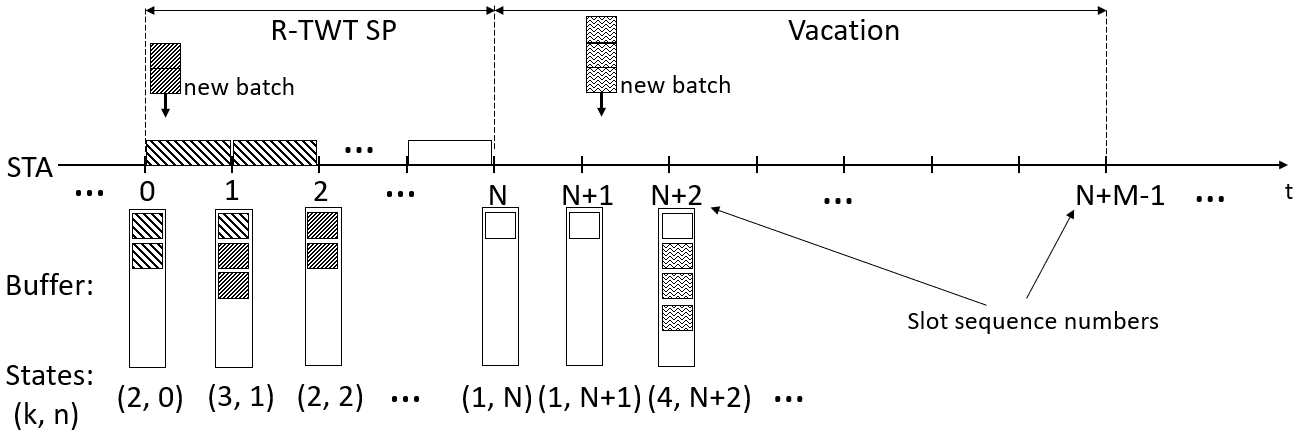}}
	\vspace{-0.3cm}
	\caption{Definition of Markov chain states.}
	\vspace{-0.6cm}
	\label{fig:markov_states}
\end{figure}

Let us consider the following Markov chain. We observe the system state at slot boundaries right after a packet is transmitted in the previous slot (if there was a packet transmission) and before a new batch of packets arrives. The system state is characterized by the following two values (cf., Figure~\ref{fig:markov_states}).

\begin{enumerate}
	\item Total number of packets in the queue $k \in [0, K]$.
	\item Slot sequence number $n \in [0, N+M-1]$, where $[0, N-1]$ correspond to SP, while $[N, N+M-1]$ --- to vacation. 
\end{enumerate}

Below we consider all possible transitions $(k, n) \rightarrow (k', n')$.

\begin{enumerate}
	\item During the \textit{vacation} ($n \in [N, N+M-1]$):
	\vspace{-0.1cm}
	\begin{enumerate}
		\item If $0$ batches arrive, or an incoming batch does not fit into the buffer, then the number of packets does not change. If the current slot has sequence number $n$, then the next one is $n+1$, unless the special case $n=N+M-1$ where the next slot is $0$.
		\vspace{-0.1cm}
		\begin{align*}
			Pr\{(k, n) \rightarrow (k, n+1)\} = \hat{b}_0 + \sum\limits_{r=K-k+1}^{R} \hat{b}_{r}.
		\end{align*}
		From here on, we assume the sums equal zero if the upper limit is less than the lower one.
		\item If a batch of size $r$ arrives and fits into the buffer (i.e., $1 \leq r \leq K-k$), then the number of packets in the buffer increases by $r$.
		\vspace{-0.1cm}
		\begin{align*}
			\Pr\{(k, n) \rightarrow (k+r, n+1)\} = \hat{b}_{r}.
		\end{align*}
	\end{enumerate}
	\item During the \textit{R-TWT SP} ($n \in [0, N-1]$):
	\begin{enumerate}
		\item If $0$ batches arrive, or an incoming batch does not fit into the buffer, then the number of packets decreases by $1$ that is transmitted during the considered slot. There is a special case when the queue is empty, then the number of packets will remain $0$ because we have nothing to transmit.
		\begin{align*}
			\hspace*{-0.5cm}\begin{cases}
			\Pr\{(0, n) \rightarrow (0, n+1)\} = \hat{b}_0+\sum\limits_{r=K+1}^{R} \hat{b}_{r};\\
			\Pr\{(k, n) \rightarrow (k-1, n+1)\} = \hat{b}_0+\sum\limits_{r=K-k+1}^{R} \hat{b}_{r},\\ 
			\hspace*{3.5cm}k \geq \max (1, K+1-R).
			\end{cases}
		\end{align*}
		\item If the batch of size $r$ arrives and fits into the buffer, then the number of packets in the buffer increases by $r-1$, because we have to exclude one packet being transmitted in the considered slot. Thus, we can write:\vspace{0.1cm}
		\hspace*{-0.2cm}$\Pr\{(k, n) \rightarrow (k+r-1, n+1)\} = \hat{b}_r, 1 \leq r \leq K-k$.
	\end{enumerate}
\end{enumerate}

The stationary distribution $\boldsymbol{p} \equiv \{p_{k, n}\}$ can be found by solving the following linear system:
\begin{align}
	\begin{cases}
			p_{k', n'} = \sum\limits_{k, n} p_{k, n} Pr\{(k, n) \rightarrow (k', n')\},\\
			\sum\limits_{k, n} p_{k, n} = 1.
	\end{cases}
\end{align}

\subsection{Packet delay distribution}

We determine the packet delay as the time between the batch arrival and the end of the transmission of the last packet from this batch. To compute the packet delay distribution, we consider only those slots where successful batches arrive. The probability that a slot belongs to state $(n, k)$ and a successful batch of size $r$ arrives can be calculated as follows:
\begin{equation}
	\label{eq:p_knr}
	\tilde{p}^{k,n}_{r} = \frac{p_{k, n} b_{r}}{b (1 - b_{fail})}.
\end{equation}

Consider a slot $n$ that belongs to a vacation, i.e., $N \leq n \leq N+M-1$. If a batch of size $r$ arrives during this slot, then the delay (in slots) for this batch will consist of the three parts.
\begin{enumerate}
	\item Remaining time of the vacation $N + M - n$ slots.
	\item Transmission time $k + r$ slots of all packets in the queue, i.e., those present before the batch arrived and the packets of the batch itself.
	\item Duration of all vacations between the R-TWT SPs, which is estimated as $D_{V}(k+r)$, where $D_V(q)$ is the number of vacation slots if a batch arrives in a queue of size $q$:
	\begin{equation}
		\label{eq:D_V_q}
		D_{V}(q) = \begin{cases}
		\lfloor \frac{q}{N} \rfloor M, \text{if}\; q \bmod N \neq 0,\\
		(\lfloor \frac{q}{N} \rfloor - 1) M, \text{if}\; q \bmod N = 0.
		\end{cases}
	\end{equation}
\end{enumerate}

Thus, the total delay for this case equals:
\begin{equation}
	\label{eq:wait_vac}
	D^{k, n}_{r} = (N+M-n) + (k+r) + D_{V}(k+r), N \leq n \leq N+M-1
\end{equation}

Now consider a slot $0 \leq n \leq N-1$ that belongs to R-TWT SP. The difference between this case and the previous one is that packets can already be transmitted within the slot $n$ and further during the remaining slots of the R-TWT SP, i.e., there is no remaining vacation time. When a batch of size $r$ arrives, the total number of packets in the queue becomes $k+r$. Then, during the R-TWT $\min(N-n, k+r)$ packets out of these $k+r$ packets will be transmitted. Hence, when the vacation starts, the remaining number of packets left in the queue before and including the last packet from the considered batch becomes $\tilde{q}^{k,n}_{r} = k+r-\min(N-n, k+r)$. Using Equation~\eqref{eq:D_V_q}, we can write the total delay for the batch as follows:
\begin{align}
	\label{eq:wait_rtwt}
	\hspace*{-0.3cm}D^{k, n}_{r} &= (k+r) + H(\tilde{q}^{k,n}_{r}) + D_{V}(\tilde{q}^{k,n}_{r}), 0 \leq n \leq N-1,
\end{align}
where $H(x)$ is the Heaviside step function.

Using Equations~\eqref{eq:p_knr},~\eqref{eq:wait_vac} and~\eqref{eq:wait_rtwt}, we can write the average packet delay:
\begin{equation}
\mathbb{E}[D] = \sum\limits_{k,n,r} \tilde{p}^{k,n}_{r} D^{k, n}_{r}
\end{equation}
\vspace{-0.3cm}

Note that in our model we approximate the continuous packet delay distribution with the discrete one, since we measure the packet delay in the integer number of slots, i.e., as a multiple of slot duration. We can calculate the probability that the packet delay equals $d$ slots as follows:
\begin{equation}
	\label{eq:wait_distr}
	\Pr\{D = d\} = \sum\limits_{k,n,r} \tilde{p}^{k, n}_{r} \; \forall (k, n, r) : D^{k, n}_{r} = d
\end{equation}

Equation~\eqref{eq:wait_distr} gives the probability distribution of the packet delay. Using this distribution, it is straightforward to calculate the moments of the distribution, variance, standard deviation (jitter), cumulative distribution function and any percentile.

\vspace{-0.25cm}
\subsection{Model limitations}
\vspace{-0.2cm}
\label{subsec:model_lims}

In this section, we discuss the limitations of the developed model that should be considered before its application.

\begin{itemize}
	\item The model relies on the Poisson arrival approximation, which is a common assumption and holds in many cases. For scenarios with other traffic generation patterns, e.g., periodic, the accuracy of the model will degrade.
	\item Time discretization simplifies the derivation of delay probability distribution, but requires that the condition $S \ll \frac{1}{\lambda}$ holds. Hence, the model accuracy will degrade in case of large packets or high traffic intensity (cf. Fig.~\ref{fig:validation3}).
	\item We consider simplified error model with constant error probability. Note that under dynamic conditions, e.g., for high mobility, the probability can vary with time.
	\item The model allows approximating high percentiles of delay, but the accuracy of the model is limited. We study the accuracy of the mode in Section~\ref{subsec:validation}.
	\item Experimental studies~\cite{barannikov2023false, liu2023first} reveal implementation issues in current commercial-off-the-shelf devices, which can affect R-TWT performance in heterogeneous networks, where Wi-Fi 7 devices coexist with ones of older Wi-Fi generations. These issues are left out of scope.
\end{itemize}

Although the aforementioned limits the applicability of the model in some scenarios, we strongly believe that it provides a powerful instrument for optimal R-TWT parameters selection, which significantly improves network performance in many practical cases (see Section~\ref{subsec:perf_eval}).

\vspace{-0.2cm}
\section{Numerical results}
\vspace{-0.1cm}
\label{sec:numerical_results}

\subsection{Model validation}
%\vspace{-0.1cm}
\label{subsec:validation}

For model validation, we consider a single Poisson flow that uses dedicated R-TWT SPs to transmit data. We list all scenario parameter values common for all experiments, unless otherwise stated, in Table~\ref{table:scenario_params}. 

\begin{table}[t]
	\centering
	\caption{\label{table:scenario_params} Scenario parameters.}
	\vspace{-0.2cm}
	\begin{tabular}{|p{0.3\textwidth}|p{0.1\textwidth}|}
		\hline
		Average inter-packet interval, $\frac{1}{\lambda}$ & $16$ms \\
		Bandwidth & $20$ MHz \\
		Buffer size, $K$ & 20 packets \\
		Error probability, $p_{err}$ & $10\%$ \\
		MCS index & 5 \\
		Number of allowed transmission attempts, $R$ & $\{1, 3\}$ \\
		Packet size & $200$ bytes \\
		Packet transmission duration, $S$ & $114.4$\textmu s \\
		\hline
	\end{tabular}
	\vspace{-0.4cm}
\end{table}

We estimate the following $4$ performance indicators:
\vspace{-0.1cm}
\begin{itemize}
	\item \textbf{average delay} $\mathbb{E}[D]$;
	\item \textbf{jitter}, i.e., delay standard deviation $\sqrt{\mathbb{E}[\left(D - \mathbb{E}[D]\right)^2]}$;
	\item \textbf{packet loss probability}, which is the probability of failure for all $R$ packet transmission attempts;
	\item \textbf{99.9\% delay percentile}, which is the minimal delay $\tilde{D}$, for which the equation $\Pr\{D < \tilde{D}\} \geq 0.999$ holds.
\end{itemize}

We estimate the described above values for different R-TWT parameters. In each experiment, we compare the results obtained with the analytical model (cf., Section~\ref{sec:math_model}) and with simulation. For that, we use an event-driven custom simulation program written in the C++ programming language~\cite{simgithub}.

\begin{figure}[t]
	\centering
	\begin{minipage}{0.4\linewidth}
		\center{\includegraphics[width=\linewidth]{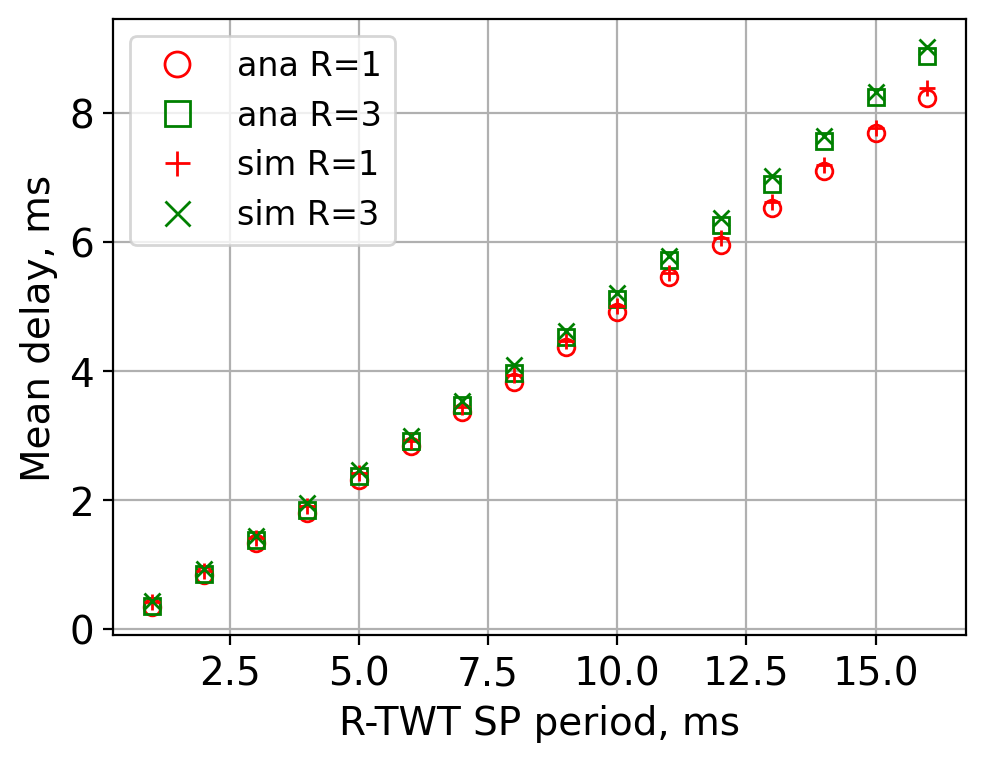} \\ 	\vspace{-0.2cm}\scriptsize{(a)}}
	\end{minipage}
	\begin{minipage}{0.4\linewidth}
		\center{\includegraphics[width=\linewidth]{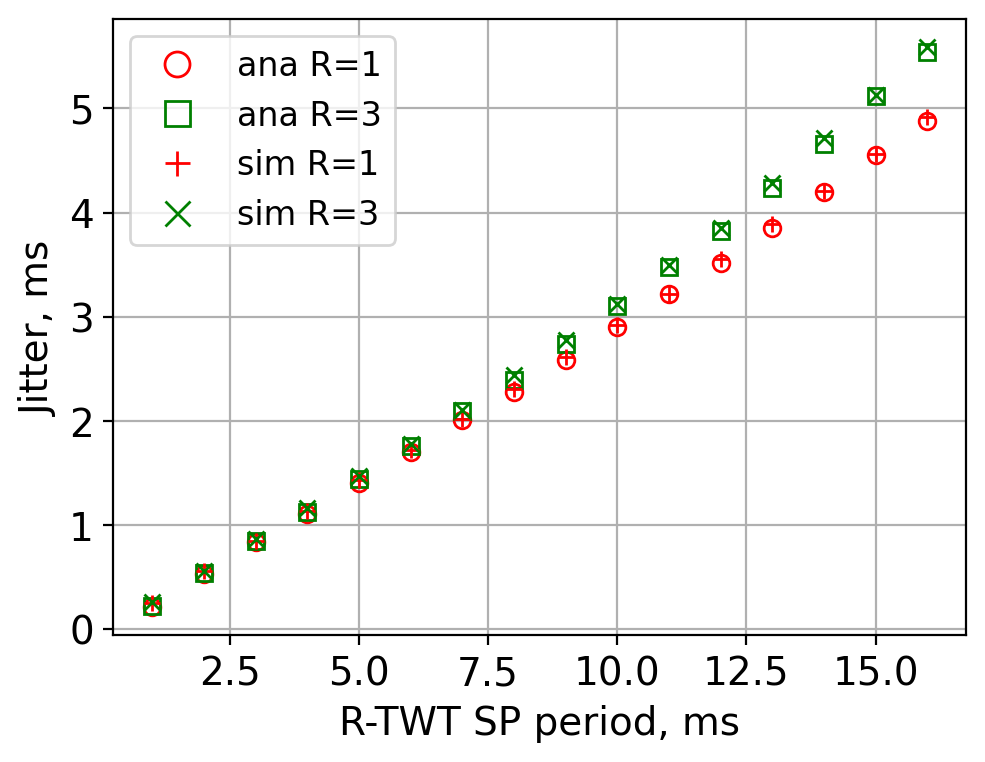} \\ 	\vspace{-0.2cm}\scriptsize{(b)}}
	\end{minipage}

	\begin{minipage}{0.4\linewidth}
		\center{\includegraphics[width=\linewidth]{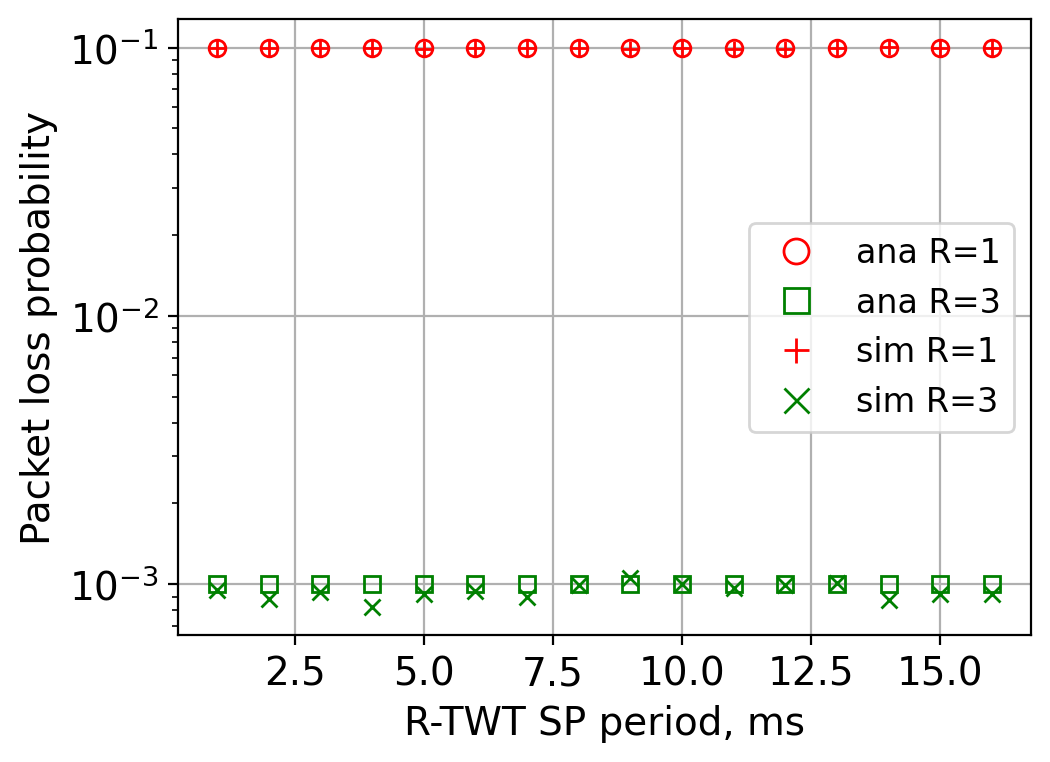} \\ 	
		\vspace{-0.2cm}\scriptsize{(c)}}
	\end{minipage}
	\begin{minipage}{0.4\linewidth}
		\center{\includegraphics[width=\linewidth]{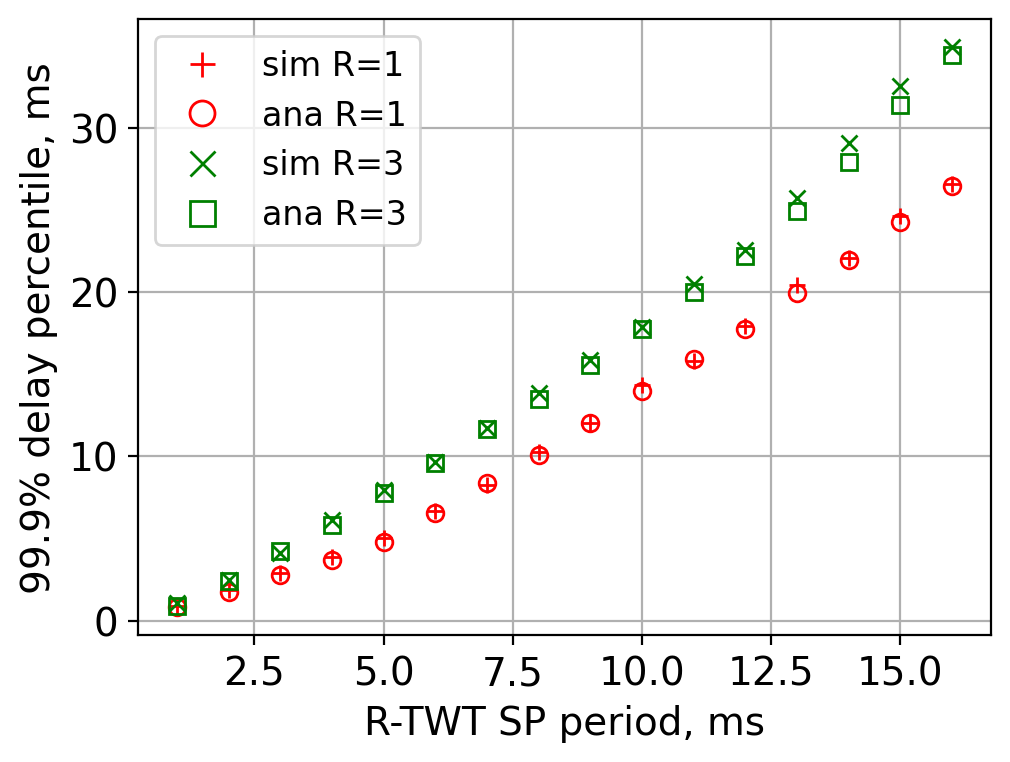} \\ 	\vspace{-0.2cm}\scriptsize{(d)}}
	\end{minipage}
	
	\vspace{-0.1cm}
	\caption{\label{fig:validation1} Model validation: R-TWT period variation. Figures: (a) average delay, (b) jitter, (c) packet loss probability, (d) 99.9\% delay percentile.}
	\vspace{-0.6cm}
\end{figure}

In the first experiment, we fix the duration of the R-TWT SP to $N=3$ slots, while varying the R-TWT period $T = \{1, \dots, 16\}$ms. Figure~\ref{fig:validation1} shows the performance indicators as functions of the R-TWT period. The label \textit{``ana''} corresponds to the analytical model, while \textit{``sim''} corresponds to simulation. We can see that the curves obtained with the analytical model fit the simulation results. For example, the maximum error for the $99.9\%$ delay percentile does not exceed $1.5$ms. The average delay, jitter and $99.9\%$ delay percentile increase with the R-TWT SP period. When $R=1$, delays are lower compared to $R=3$, because for $R=1$ there are no packet retransmissions, and thus each packet spends less time in the queue. For example, to maintain $99.9\%$ percentile at the $10$ms level, the R-TWT period should be set to $6$ms for $R=3$, while for $R=1$ a period of $8$ms is enough. On the other hand, the packet loss probability significantly decreases with increasing the number of transmission attempts while being independent of the R-TWT SP period. Specifically, each additional transmission attempt lowers the packet loss probability by one order of magnitude, because $p_{err}=0.1$.

\begin{figure}[t]
	\centering
	\begin{minipage}{0.4\linewidth}
		\center{\includegraphics[width=\linewidth]{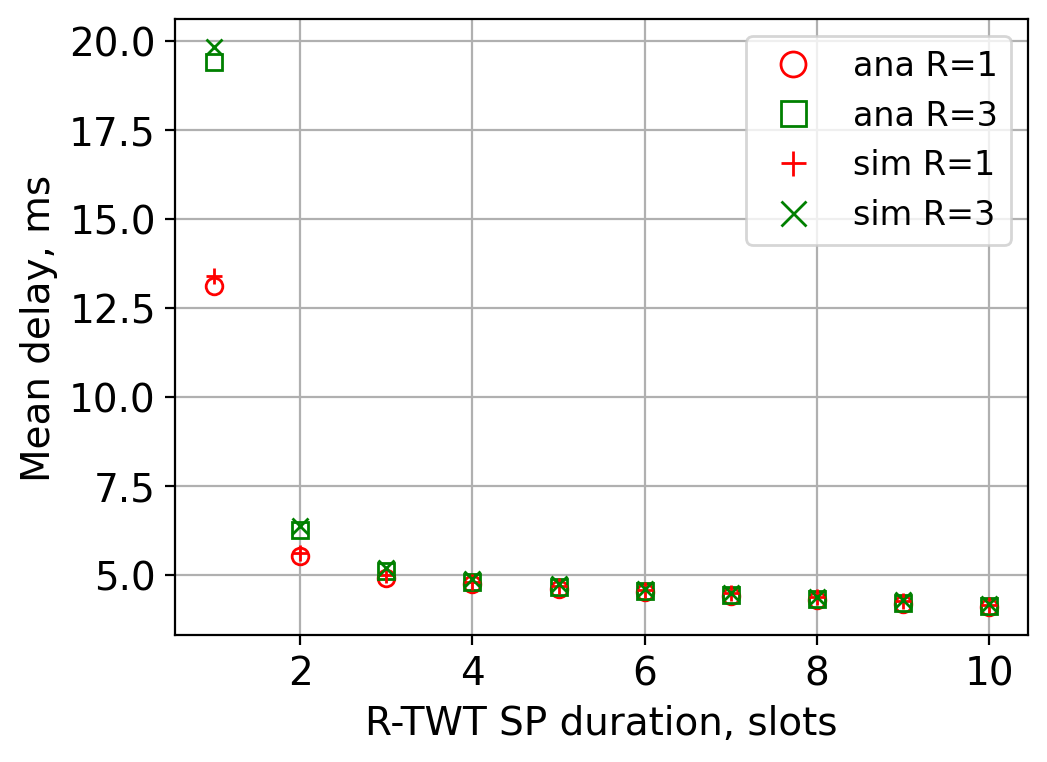} \\ 	\vspace{-0.2cm}\scriptsize{(a)}}
	\end{minipage}
	\begin{minipage}{0.4\linewidth}
		\center{\includegraphics[width=\linewidth]{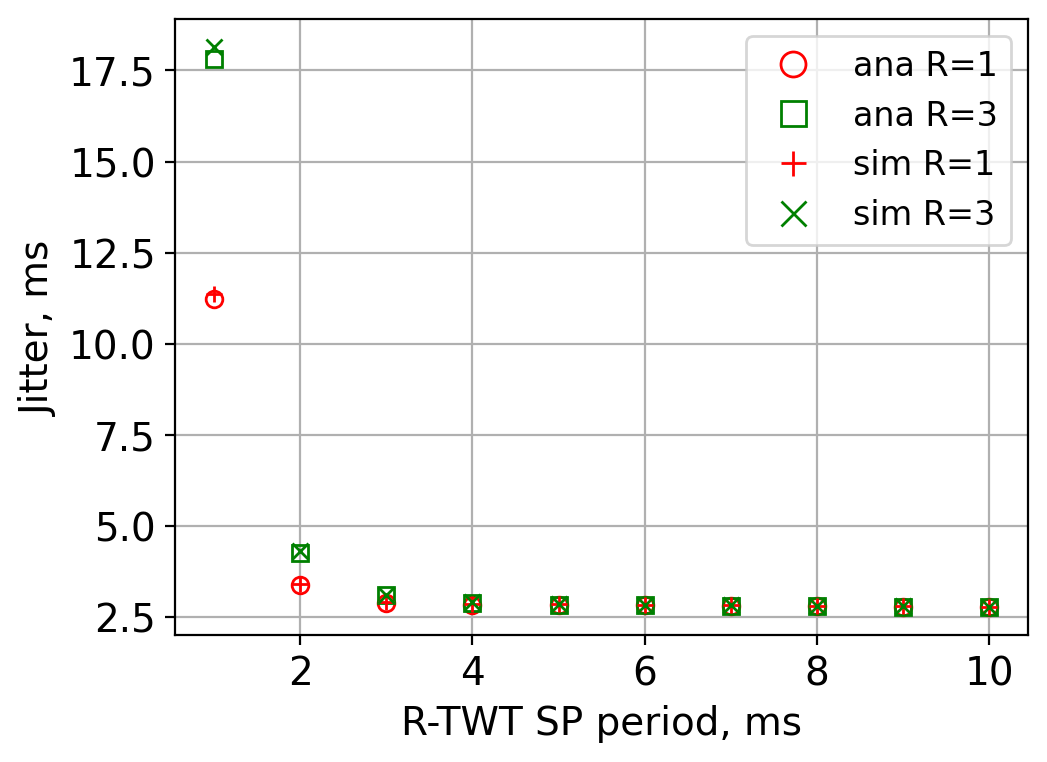} \\ 	\vspace{-0.2cm}\scriptsize{(b)}}
	\end{minipage}
	
	\begin{minipage}{0.4\linewidth}
		\center{\includegraphics[width=\linewidth]{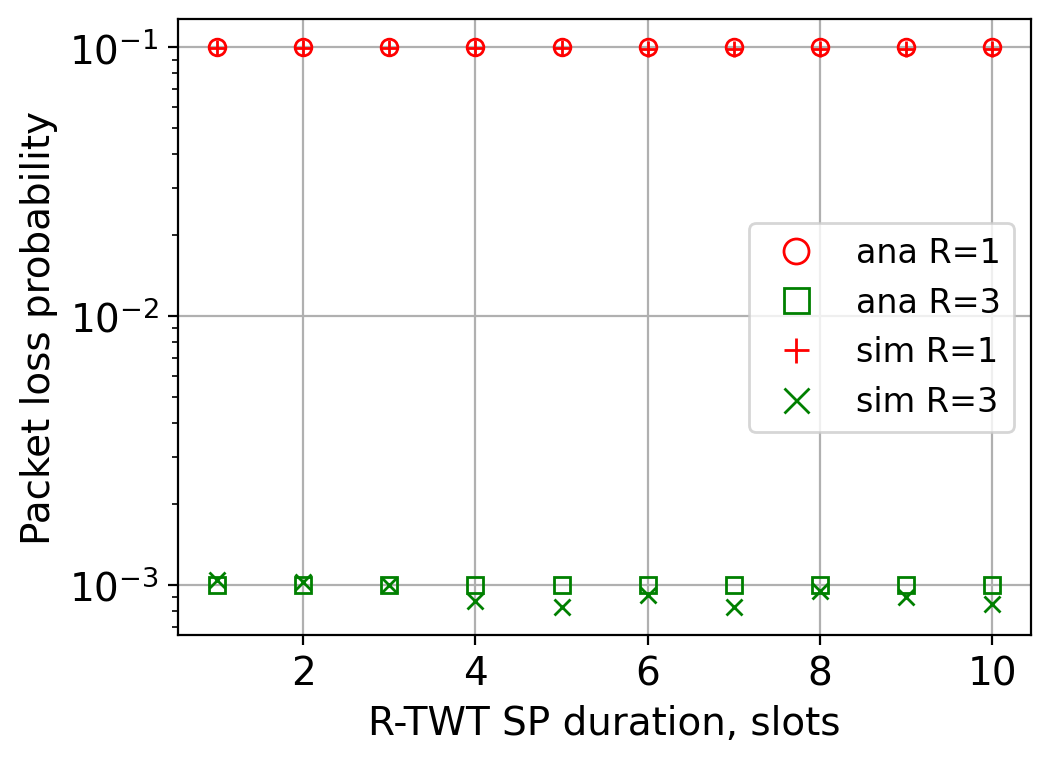} \\ 	
		\vspace{-0.2cm}\scriptsize{(c)}}
	\end{minipage}
	\begin{minipage}{0.4\linewidth}
		\center{\includegraphics[width=\linewidth]{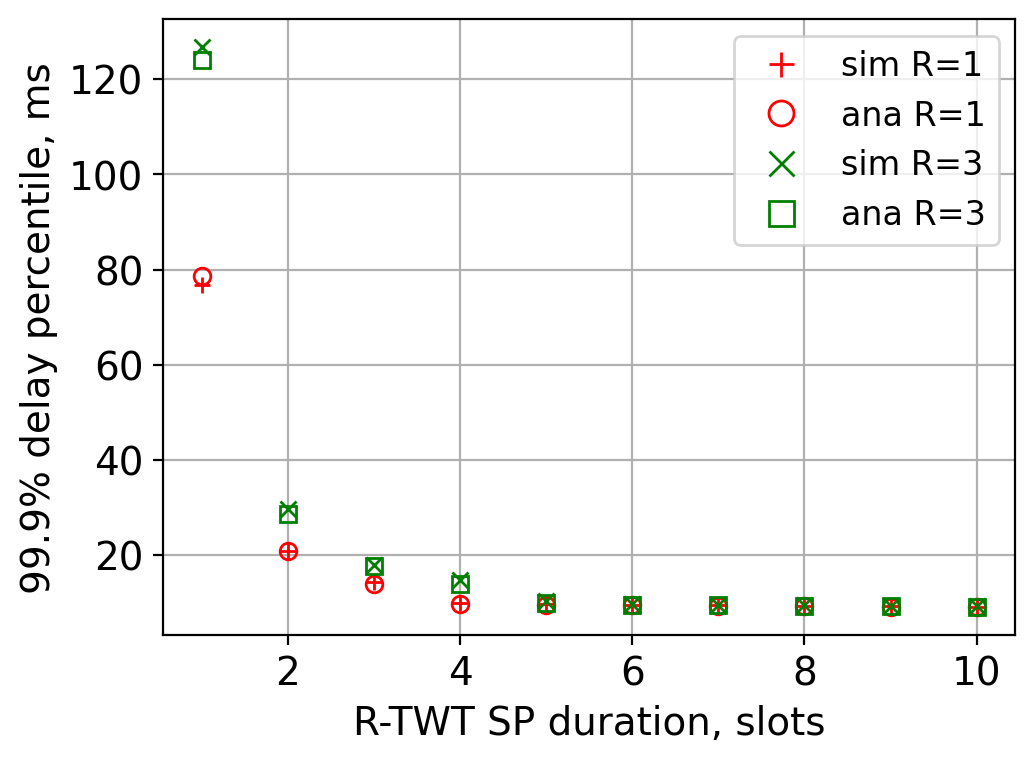} \\ 	\vspace{-0.2cm}\scriptsize{(d)}}
	\end{minipage}
	
	\vspace{-0.1cm}
	\caption{\label{fig:validation2} Model validation: R-TWT SP duration variation. Figures: (a) average delay, (b) jitter, (c) packet loss probability, (d) 99.9\% delay percentile.}
	\vspace{-0.4cm}
\end{figure}

In the second experiment, we fix the R-TWT period to $T = 10$ms and vary the R-TWT SP duration $N=\{1, \dots, 10\}$ (cf., Figure~\ref{fig:validation2}). Similar to the previous experiment, the results from the analytical model fit the simulation results well (the error for $99.9\%$ delay percentile does not exceed $3$ms). Packet delays decrease when the R-TWT SP increases, which is expected behavior because the station has more opportunities to transmit its packets without waiting for the next R-TWT SP after a long vacation period. We can see that increasing the R-TWT SP duration initially significantly decreases the delay, but after $N = 5$ further increase does not provide any noticeable effect. In particular, the delay percentile drops from tens of milliseconds to approximately $9$ms and does not decrease anymore. The reason for such behavior is that the load offered by the RTA flow ($1$ packet per $16$ms) results in a low probability of the queue containing more than 5 packets.

\begin{figure}[t]
	\centerline{\includegraphics[width=0.4\linewidth]{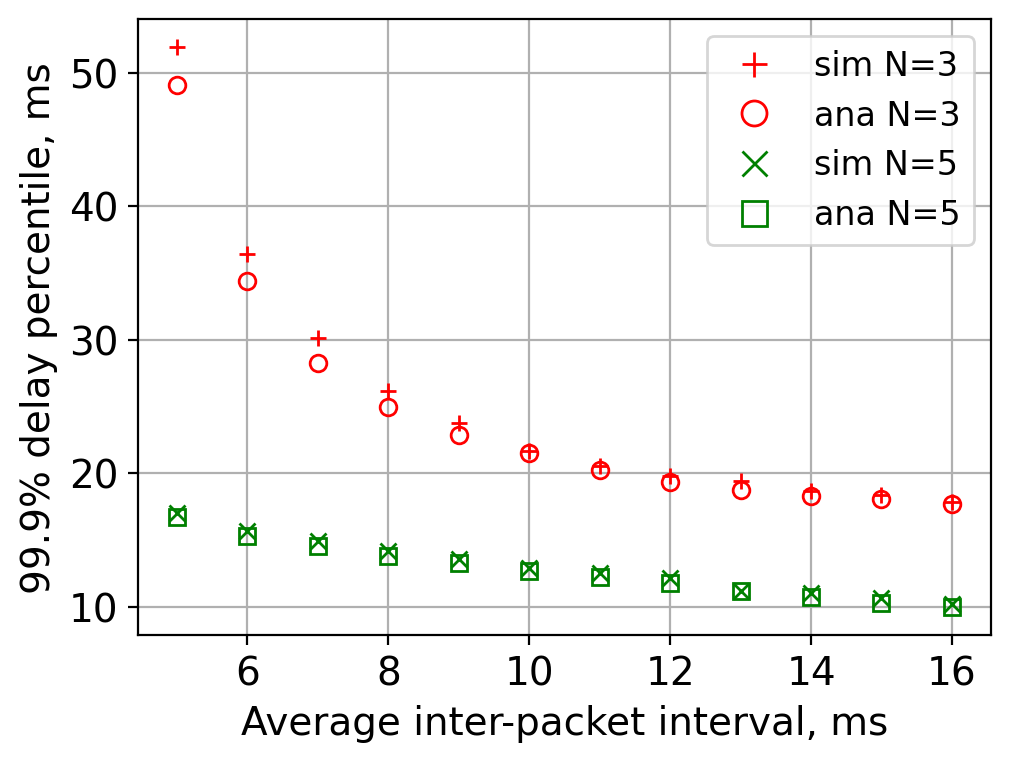}}
	\vspace{-0.2cm}
	\caption{Model validation: variation of load.}
	\vspace{-0.6cm}
	\label{fig:validation3}
\end{figure}

To investigate how the delay varies with the load, we conduct the following experiment. We fix both the R-TWT period and number of allowed transmission attempts to $T = 10$ms and $R = 3$ respectively, and vary the average inter-packet interval $\frac{1}{\lambda} = \{5, \dots, 16\}$ms and R-TWT SP duration $N=\{3, 5\}$ packets (cf., Figure~\ref{fig:validation3}). When the inter-packet interval increases the offered load decreases, and thus the number of packets in the queue during the experiment also decreases. Hence, we see that delay percentile decreases too. Besides, we can see that even for $N=5$ packets the delay percentile is not constant, because the probability of having more than $5$ packets in the queue increases with the load. Note that the model becomes less accurate for higher load. This happens due to the model assumption that $S \ll \frac{1}{\lambda}$, which holds less strictly when $\lambda$ increases. However, even for $T = 5$ms the accuracy is very good, with an error of around $5$\%.

\vspace{-0.3cm}
\subsection{Performance evaluation}
\vspace{-0.1cm}
\label{subsec:perf_eval}

\begin{figure}[t]
	\centering
	\begin{minipage}{0.4\linewidth}
		\center{\includegraphics[width=\linewidth]{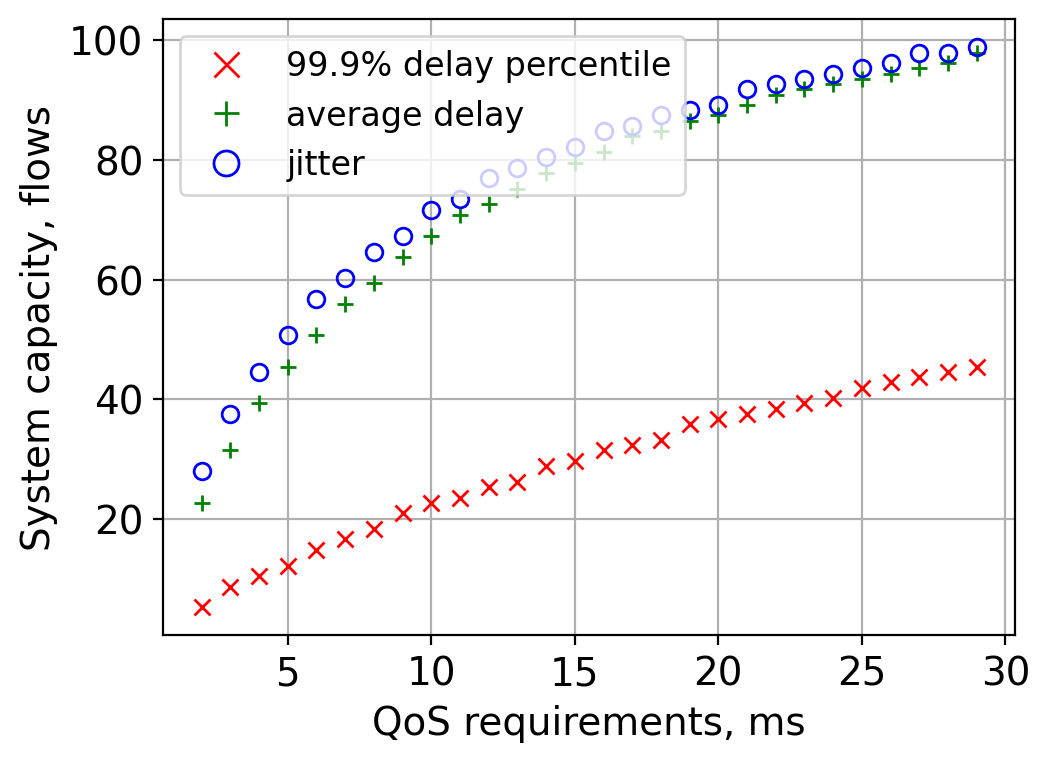} 
		\\ \vspace{-0.2cm}\scriptsize{(a)}}
	\end{minipage}
	\begin{minipage}{0.4\linewidth}
		\center{\includegraphics[width=\linewidth]{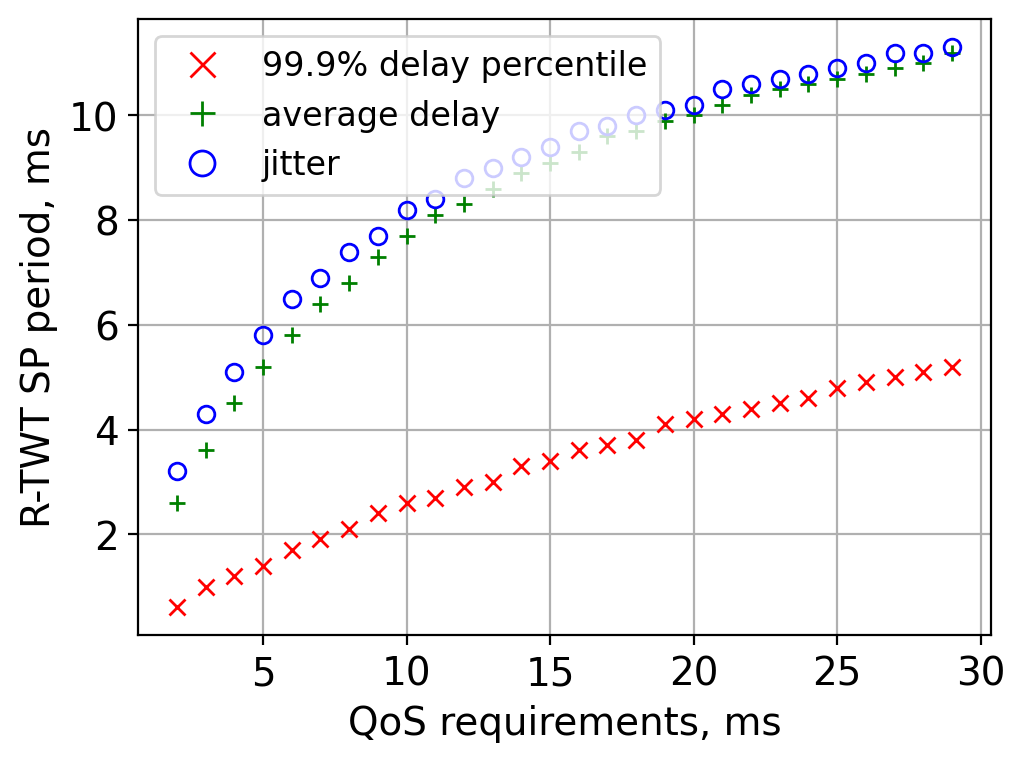} 
		\\ \vspace{-0.2cm}\scriptsize{(b)}}
	\end{minipage}
	\vspace{-0.1cm}
	\caption{\label{fig:evaluation} Performance evaluation: (a) system capacity, (b) selected period.}
	\vspace{-0.7cm}
\end{figure}

For performance evaluation, we fix the number of transmission attempts $R$ to $3$, which corresponds to a loss probability $10^{-3}$. All other parameters are the same as listed in Table~\ref{table:scenario_params}. We vary the QoS target value from $1$ to $30$ ms for three RTA performance indicators: 1) $99.9\%$ delay percentile; 2) average delay; 3) jitter. For each target value, we find the optimal R-TWT period and SP duration, i.e., the pair of values that provides the highest system capacity $C = \frac{T}{N\cdot S}$ while ensuring that the performance indicator stays below the target value. To find the optimal values, we use the analytical model and an exhaustive search algorithm. Specifically, we vary the R-TWT period $T$ between $0.5$ms and $16$ms with a step of $0.1$ms, and the SP duration $N$ between $1$ and $5$ slots. Figure~\ref{fig:evaluation} shows the system capacity and the optimal R-TWT period as a function of the target QoS value. We do not show the figure for selected SP duration, because it equals $1$ for all target values. This means that transmitting more packets within the same R-TWT SP for the considered RTA flow has less benefits compared to decreasing the R-TWT period, because the load offered by the RTA flow is relatively low and does not cause packet queuing.

Figure~\ref{fig:evaluation} shows that delay percentile restricts the selection of the R-TWT period more significantly than jitter and average delay. For example, if the target value for delay percentile is $20$ms, then the AP has to select a $4$ms R-TWT period, which results in supporting $40$ simultaneous flows. With the same R-TWT period we provide less than $3$ms average delay and jitter. If we consider even stricter requirements, e.g., $5$ms delay percentile, then the system capacity is less than $10$ flows, assuming that the whole channel airtime is given to RTA flows. 

\vspace{-0.3cm}
\subsection{Discussion and future research directions}
\vspace{-0.1cm}

In real scenarios, other types of traffic might coexist with RTA flows, thus the system capacity will be lower than estimated in Section~\ref{subsec:perf_eval}. Besides, the allocation of dedicated R-TWT SPs significantly limits the system scalability. To address these issues, several approaches deserve attention. First, RTA flows can benefit from participation in random channel access in between R-TWT SPs. Then packets will experience lower delay, and the AP can assign less dedicated resources to each flow. Second, Orthogonal Frequency-Division Multiple Access (OFDMA), which is another promising technology introduced in recent Wi-Fi generations, can be combined with R-TWT, thus allowing RTA flows to transmit simultaneously in the same SP in different resource units without contention. These two approaches are considered for future research.

In addition, real scenarios often comprise fluctuating network conditions due to mobility of the stations, or interference from neighboring networks. Improving the model for tackling such scenarios will be also considered in future studies.

\vspace{-0.2cm}
\section{Conclusion}
\label{sec:conclusion}
\vspace{-0.1cm}

In this paper, we have proposed a novel approach for mathematical modeling of R-TWT for RTA applications in Wi-Fi 7 networks. Specifically, we replace the continuous-time system with a slotted one, where the size of one slot equals the packet transmission time, and multiple transmission attempts for individual packets are modeled as batch arrivals. Our approach significantly lowers the computational complexity, allowing it to be used for real-time optimization of R-TWT parameters. For a given set of parameters, the developed analytical model allows estimating the delay distribution and the packet loss probability. Using the delay distribution, the AP can further estimate the performance indicators, such as average delay, jitter and delay percentile. The model provides an instrument for the AP to select the optimal R-TWT parameters that satisfy RTA QoS requirements while maximizing the system capacity. In our future research, we plan to extend the model for scenarios with mixed R-TWT and random channel access, including the usage of the OFDMA technology.

\vspace{-0.2cm}
\bibliographystyle{IEEEtran}
\bibliography{biblio}

% Generated by IEEEtran.bst, version: 1.12 (2007/01/11)
\begin{thebibliography}{10}
\providecommand{\url}[1]{#1}
\csname url@samestyle\endcsname
\providecommand{\newblock}{\relax}
\providecommand{\bibinfo}[2]{#2}
\providecommand{\BIBentrySTDinterwordspacing}{\spaceskip=0pt\relax}
\providecommand{\BIBentryALTinterwordstretchfactor}{4}
\providecommand{\BIBentryALTinterwordspacing}{\spaceskip=\fontdimen2\font plus
\BIBentryALTinterwordstretchfactor\fontdimen3\font minus
  \fontdimen4\font\relax}
\providecommand{\BIBforeignlanguage}[2]{{%
\expandafter\ifx\csname l@#1\endcsname\relax
\typeout{** WARNING: IEEEtran.bst: No hyphenation pattern has been}%
\typeout{** loaded for the language `#1'. Using the pattern for}%
\typeout{** the default language instead.}%
\else
\language=\csname l@#1\endcsname
\fi
#2}}
\providecommand{\BIBdecl}{\relax}
\BIBdecl

\bibitem{rta2018report}
{IEEE 802.11 Real Time Applications TIG Report}.

\bibitem{khorov2020current}
E.~Khorov \emph{et~al.}, ``{Current Status and Directions of IEEE 802.11 be,
  the Future Wi-Fi 7},'' \emph{IEEE Access}, vol.~8, pp. 88\,664--88\,688,
  2020.

\bibitem{ieee80211be}
``{IEEE P802.11be/D2.1, Draft Standard for Information Technology.}''

\bibitem{tian2021wi}
L.~Tian \emph{et~al.}, ``{Wi-Fi HaLow for the Internet of Things: An Up-to-Date
  Survey on IEEE 802.11 ah Research},'' \emph{J. Netw. Comput. Appl.}, 2021.

\bibitem{nurchis2019target}
M.~Nurchis \emph{et~al.}, ``{Target wake time: Scheduled Access in IEEE 802.11
  ax WLANs},'' \emph{IEEE Wirel. Commun.}, vol.~26, no.~2, pp. 142--150, 2019.

\bibitem{memon2021survey}
S.~K. Memon \emph{et~al.}, ``{A Survey on 802.11 MAC Industrial Standards,
  Architecture, Security \& Supporting Emergency Traffic: Future Directions},''
  \emph{J. Ind. Inf. Integr.}, vol.~24, p. 100225, 2021.

\bibitem{isolani2021support}
P.~H. Isolani \emph{et~al.}, ``{Support for 5G Mission-Critical Applications in
  Software-Defined IEEE 802.11 Networks},'' \emph{Sensors}, vol.~21, 2021.

\bibitem{richart2020slicing}
M.~Richart \emph{et~al.}, ``{Slicing with Guaranteed Quality of Service in WiFi
  Networks},'' \emph{IEEE Trans. Netw. Service Manag.}, vol.~17, no.~3, 2020.

\bibitem{chemrov2022smart}
K.~Chemrov \emph{et~al.}, ``{Smart Preliminary Channel Access to Support
  Real-Time Traffic in Wi-Fi Networks},'' \emph{Fut. Int.}, vol.~14, no.~10, p.
  296, 2022.

\bibitem{xia2023hybrid}
W.~Xia \emph{et~al.}, ``{Hybrid Channel Access Towards Real-Time Applications
  in Healthcare},'' in \emph{Proc. of IEEE ICC WS21}, 2023, pp. 1932--1937.

\bibitem{yang2018energy}
H.~Yang \emph{et~al.}, ``{On Energy Saving in IEEE 802.11 ax},'' \emph{IEEE
  Access}, vol.~6, pp. 47\,546--47\,556, 2018.

\bibitem{santi2019accurate}
S.~Santi \emph{et~al.}, ``{Accurate Energy Modeling and Characterization of
  IEEE 802.11 ah RAW and TWT},'' \emph{Sensors}, vol.~19, no.~11, p. 2614,
  2019.

\bibitem{stepanova2020joint}
E.~Stepanova \emph{et~al.}, ``{On the Joint Usage of Target Wake Time and
  802.11 ba Wake-Up Radio},'' \emph{IEEE Access}, vol.~8, p. 221061, 2020.

\bibitem{guo2020performance}
X.~Guo \emph{et~al.}, ``{Performance Evaluation of the Networks with Wi-Fi
  based TDMA Coexisting with CSMA/CA},'' \emph{Wirel. Pers. Commun.}, 2020.

\bibitem{khorov2020modeling}
E.~Khorov \emph{et~al.}, ``{Modeling of Real-Time Multimedia Streaming in Wi-Fi
  Networks with Periodic Reservations},'' \emph{IEEE Access}, vol.~8, 2020.

\bibitem{kleinrock1975queueing}
L.~Kleinrock, \emph{{Queueing Systems. Volume 1: Theory}}.\hskip 1em plus 0.5em
  minus 0.4em\relax Wiley-Interscience.

\bibitem{lee1989m}
T.~T. Lee, ``{M/G/1/N queue with vacation time and limited service
  discipline},'' \emph{Performance Evaluation}, vol.~9, no.~3, pp. 181--190,
  1989.

\bibitem{de1982improved}
F.~R. De~Hoog \emph{et~al.}, ``{An Improved Method for Numerical Inversion of
  Laplace Transforms},'' \emph{J Sci Comput.}, vol.~3, no.~3, pp. 357--366,
  1982.

\bibitem{barannikov2023false}
A.~Barannikov \emph{et~al.}, ``{False Protection of Real-Time Traffic with
  Quieting in Heterogeneous Wi-Fi 7 Networks: An Experimental Study},''
  \emph{Sensors}, vol.~23, no.~21, p. 8927, 2023.

\bibitem{liu2023first}
R.~Liu \emph{et~al.}, ``{A First Look at Wi-Fi 6 in Action: Throughput,
  Latency, Energy Efficiency, and Security},'' \emph{ACM POMACS}, vol.~7, pp.
  1--25, 2023.

\bibitem{simgithub}
\BIBentryALTinterwordspacing
{Event-Driven Custom Simulator for Modeling of R-TWT}. [Online]. Available:
  \url{https://github.com/imec-idlab/R-TWT-sim-public}
\BIBentrySTDinterwordspacing

\end{thebibliography}

\end{document}